\documentclass[12pt]{article}
\usepackage{amsfonts}

\newcommand{\sect}[1]{\setcounter{equation}{0}\section{#1}}
\newcommand{\subsect}[1]{\subsection{#1}}

\def\be{\begin{equation}}
\def\ee{\end{equation}}
\def\bea{\begin{eqnarray}}
\def\eea{\end{eqnarray}}

\def\1{\'{\i}}                           

\def\expandj{J'}
\def\expandp{P'}
\def\expandh{H'}
\def\expandk{K'}

\def\jja{{\cal J}_1}
\def\ppa{{\cal C}_1}
\def\jjb{{\cal J}_2}
\def\ppb{{\cal C}_2}
\def\jjj{{\cal J}}
\def\ppp{{\cal C}}

\def\expandcasa{{\cal C}'_1}
\def\expandcasb{{\cal C}'_2}

\def\kk{\kappa}

\def\k{\omega}

\def\ext{m}

\def\p{t}
\def\h{k}
\def\AA{\alpha}
 
\def\>#1{{\bf #1}}                 
\def\back{\!\!\!\!\!\!}
\def\Back{\back\back\back\back\back}

\begin{document}

\thispagestyle{empty}

\ 
\vspace{2cm}

\begin{center}
{\LARGE{\bf{2+1  Kinematical expansions:}}}
 
{\LARGE{\bf{from Galilei to de Sitter algebras}}}
\end{center}

\bigskip\bigskip

\begin{center}
Francisco J. Herranz$^\dagger$ 
and Mariano Santander$^\ddagger$
\end{center}

\begin{center}
{\it $^\dagger$ Departamento de F\'{\i}sica, Escuela
Polit\'ecnica Superior\\ 
Universidad de Burgos, E--09006 Burgos, Spain}
\end{center}

\begin{center}
{\it $^{\ddagger}$ Departamento de F\'{\i}sica Te\'orica,
Facultad de Ciencias\\ Universidad de Valladolid,
E--47011 Valladolid, Spain}
\end{center}

\bigskip\bigskip\bigskip

\begin{abstract} 
Expansions of Lie algebras are the opposite process   of
contractions. Starting from a Lie algebra, the expansion
process goes to another one, non-isomorphic and less
abelian.  We propose an expansion  method based in the
Casimir  invariants of the initial and expanded algebras and
where the free parameters involved in the expansion are the
curvatures of their associated homogeneous spaces. This
method is applied for  expansions within the family of Lie
algebras of 3d spaces and (2+1)d  kinematical algebras. We
show that these expansions are classed in two types.  The
first type makes different from zero the curvature of space
or space-time (i.e., it introduces a space or universe
radius), while the other has a similar interpretation for the
curvature of the space of worldlines, which is non-positive
and equal to $-1/c^2$ in the kinematical algebras.  We get
expansions which go from Galilei to either Newton--Hooke or
Poincar\'e algebras, and from these ones to de Sitter
algebras, as well as some other examples. 
\end{abstract}

\newpage 


\sect{Introduction}

The concept of contraction of Lie algebras and  groups arose in
the study of the limit from relativistic to classical mechanics.
As it is well known,  when the velocity of light  goes  to
infinity the Poincar\'e group leads formally to the Galilei one. 
This idea, proposed and studied by In\"on\"u and Wigner \cite{IW}
appeared also in Segal \cite{Segal} and was later developed by
Saletan
\cite{Saletan}. More recently, other approachs to the study of
contractions, such  as the graded contraction theory
\cite{MonPat,MooPat} and the generalized In\"on\"u--Wigner
contractions \cite{Weimar} have been introduced. In general,
a Lie algebra  contraction starts from some Lie  algebra and
makes to vanish some non-zero structure constants giving rise to
another Lie algebra which is more   abelian than the original
one. The theory of graded contractions includes In\"on\"u--Wigner
contractions but goes beyond that and, for instance, may also
relate different real forms of  semisimple Lie algebras.

The  opposite process of a  contraction  limit is generically, and
rather imprecisely called an {\em expansion}.
One specific way to implement the  expansion idea is to replace
the generators of the initial algebra by some  functions of them;
if the new generators close a Lie algebra then we have obtained an
expansion of the original algebra \cite{Gilmore}. This kind of
process produces, so to speak, some non-zero structure constants
which were previously equal to zero,  in such a manner that the
final algebra is {\em less} abelian than the initial one. In this
approach usually the expanded algebra is realized as a subalgebra
within an irreducible representation of the universal enveloping
algebra for the initial algebra. We remark that  algebra
expansions are also called in the literature  algebra deformations
and indeed this kind of process can be seen as a `classical
deformation'. However, we will use the former name in order to
avoid confusion with quantum algebra deformations.

Unlike the study of Lie algebra contractions, the theory  of
expansions has not been systematized.  Known expansions are those
going from the inhomogeneous pseudo-orthogonal algebras
$iso(p,q)$  to the semisimple ones $so(p+1,q)$ and
similar expansions   for the unitary algebras from $iu(p,q)$ to $
u(p+1,q)$  \cite{Gilmore,Rosen}. On the other hand,  a different
procedure which allows to perform  expansions $t_{qp}(so(p)\oplus
so(q))\to so(p,q)$ or 
$t_{qp}(u(p)\oplus u(q))\to u(p,q)$ as well as their symplectic
analogous has been introduced in
\cite{WB} (see also references therein).

The set of quasi-orthogonal algebras \cite{gradedb}  appears as a
natural frame for  developing a  study of expansions, with a good
balance between generality and suitability as an adapted tool for
specific purposes. This set of algebras includes all
pseudo-orthogonal algebras as well as a large number of graded
contractions ---relative to a given maximal fine grading--- of
the simple (pseudo)-orthogonal algebras. These contractions are
however not the most general ones, but still somehow keep the
properties linked to simplicity, an important fact which makes
these algebras a natural subset among all graded contractions of
the orthogonal algebras. When turning to expansions, these
remarks should be reversed: it is true that in principle  {\em
any} Lie  algebra can be realized in the universal enveloping
algebra of a direct product of Heisenberg algebras, as Schwinger
realizations for the simple cases clearly show
\cite{BarutRaczka}. But as in the most general set  of
contractions,  it seems pertinent to restrict oneself to the
study of expansions amongst quasi-simple algebras, which should
reverse the contractions found in this family. For instance  we
would find, amongst many other expansions, the ones concerning
the kinematical algebras
\cite{BLL}: expansions  going from  Newton--Hooke to  de Sitter 
algebras, from Galilei to   Newton--Hooke or to  Poincar\'e
algebras, further to expansions from Poincar\'e to the de Sitter
or from Galilei to the two Newton-Hooke algebras. Notice that the
known transitions $iso(p,q)\to so(p+1,q)$ mentioned above include
only expansions from Poincar\'e to de Sitter algebras, but not
for the remaining quoted cases.  As far as we know, specific
possibilities for such general expansion schemes have not been
studied in some generality; see however
\cite{nieto} for the (1+1)-dimensional case.

The aim of this paper is to provide a   simple new expansion
procedure and to apply it  to the manageable but non-trivial case
of the Lie algebras of motion groups in 3d spaces. These
expansions would include all the expansions of kinematical
algebras in 2+1 dimensions and a few more, non kinematical
examples.  Thus in next section we present the structure of the
main  kinematical algebras which include three `absolute  time'
cases  (two Newton--Hooke and Galilei) and three `relative time'
ones (two de Sitter and Poincar\'e).  In section 3 we  propose an
expansion method which  is based in the Casimir invariants of the
two Lie algebras involved in the expansion. For instance, the
initial  Lie algebra may be  related to  a space-time of
curvature zero, while the expanded algebra corresponds to an
homogeneous space with constant non-zero curvature; from this
point of view we will see how the expansion process introduces
the curvature as a {\em free} parameter.  The remaining sections
of the paper are devoted to analyze in detail all the possible
kinematical expansions casted into two types: `space-time'
expansions which starting from the algebra of a flat space-time
will introduce curvature in space-time, and `speed-space'
expansions which recover a `relative time' space-time with a
finite relativistic constant $c$ (equal to the velocity of the
light), starting from the algebra of an `absolute time' (and
hence $c=\infty$) space-time.


\sect{The (2+1)-dimensional kinematical algebras}

Let us consider  an homogeneous space-time with curvature  $\kk$
and either of `absolute time' type (formally described by letting
$c\to\infty$) or `relative time' one, with relativistic constant
$c$.  Let $H$, $P_i$, $K_i$ $(i=1,2)$ and
$J$  the  generators of time translations, space  translations,
boosts and spatial rotations, respectively. The structure  of the
kinematical algebras we are going to deal with can be written
collectively in terms of two real coefficients $\k_1=\kk$ and
$\k_2=-1/c^2$ as follows 
\be
\begin{array}{lll}
[J,P_i]=\epsilon_{ij} P_j&\qquad [J,K_i]=\epsilon_{ij} K_j&
\qquad [J,H]=0\cr
[P_1,P_2]=\k_1\k_2 J&\qquad [K_1,K_2]=\k_2 J &
\qquad [P_i,K_j]=\delta_{ij}\k_2 H\cr
[H,P_i]=\k_1 K_i&\qquad [H,K_i]=-P_i &\qquad i,j=1,2
\end{array}
\label{aa}
\ee
where $\epsilon_{ij}$ is a skewsymmetric tensor such that
$\epsilon_{12}=1$, $\epsilon_{21}=-1$  and
$\epsilon_{11}=\epsilon_{22}=0$. The main reason to introduce
$\k_2$ instead of $c$ is to allow {\em positive} values,
whenever  (\ref{aa}) has not a kinematical interpretation, but
nevertheless makes perfect sense as a Lie algebra. Thus each
coefficient
$\k_i$  can take positive, negative or zero  values,  and the
commutators (\ref{aa}) give rise to nine Lie algebras, which
should be considered as different in this context. For each such
algebra a symmetric homogeneous space can be built up by taking
the quotient by the subalgebra generated by
$K_i, J$. These Lie algebras as well as the  homogeneous spaces
are displayed in table 1 according to the values of the pair
$(\k_1,\k_2)$.

The value of $\k_2$ can be thought
of as related to the signature of the metric in the homogeneous
space, which is definite positive for
$\k_2>0$ and  indefinite (hence Lorentzian  type in this 3d case)
for $\k_2<0$, with the Galilean degenerate
metric which corresponds to `absolute time'  in the
intermediate case $\k_2=0$.    Therefore, the three algebras of
the first row with  $\k_2>0$ do not allow a {\em literal\/}
interpretation in terms of a space-time, and instead they are the
Lie algebras of the motion groups of three-dimensional {\em
riemannian} spaces of constant curvature $\k_1=\kk$. Kinematical
algebras
\cite{BLL} arise whenever
$\k_2\le 0$, that is, when the boosts generate  non-compact
subgroups. The coefficient $\k_1$ is the universe curvature
$\kk$; the so called universe radius 
$R$ is related with  $\k_1$ by either $\k_1= 
\frac{1}{R^2}$ or $\k_1=  -\frac{1}{R^2}$. The relativistic
constant $c$ plays a role analogous to $R$ when
$\k_2$ is {\em negative}, $\k_2= - \frac{ 1}{{c^2}}$. 
The three algebras of the second row (NH means
Newton--Hooke) correspond to `absolute time' space-times with
$\k_2=0$ or $c=\infty$,  while those of the third row  are
associated to  `relative time'  space-times with $\k_2<0$ and a
finite value for $c$.
 
\medskip

{\bf Table 1.} 3d isometry Lie algebras and their
homogeneous spaces, including (2+1)d kinematical algebras.

\noindent\hfill
\begin{tabular}{ccccc}
\hline
$so(4)$&$\Back\longrightarrow\Back$&
$iso(3)$&$\Back\longleftarrow\Back$&
$so(3,1)$\\
$(+,+)$& &$(0,+)$& &$(-,+)$\\
3d Elliptic space& &3d Euclidean space& &3d Hyperbolic
space\\[0.2cm]
$\downarrow$& &$\downarrow$& &$\downarrow$\\[0.2cm]
$t_4(so(2)\oplus
so(2))$&$\Back\longrightarrow\Back$&
$iiso(2)$&$\Back\longleftarrow\Back$&$t_4(so(2)\oplus
so(1,1))$\\ 
$(+,0)$& &$(0,0)$& &$(-,0)$\\
Oscillating NH & &Galilean  & 
&Expanding NH \\ 
(2+1)d space-time& &(2+1)d   space-time  & 
&(2+1)d space-time\\[0.2cm]
$\uparrow$& &$\uparrow$& &
$\uparrow$\\[0.2cm]
$so(2,2)$&$\Back\longrightarrow\Back$&
$iso(2,1)$&$\Back\longleftarrow\Back$&
$so(3,1)$\\
$(+,-)$& &$(0,-)$& &$(-,-)$\\
Anti-de Sitter  & &Minkowskian &
&de Sitter  \\
  (2+1)d  space-time& & (2+1)d space-time&
& (2+1)d space-time\\[0.1cm]
\hline
\end{tabular} 
\hfill
\bigskip

These Lie  algebras have two Casimir invariants given by:
\be
\begin{array}{l}
{\cal C}_1=\k_2 H^2+P_1^2+P_2^2+
\k_1(K_1^2+ K_2^2)+ \k_1\k_2 J^2 \cr
{\cal C}_2=\k_2 H J - P_1 K_2 + P_2 K_1  
\end{array} 
\label{ab}
\ee
which in the kinematical cases $\k_2\leq 0$ correspond  to the
energy and angular momentum of a particle in the free kinematics
of the space-time corresponding to $(\k_1, \k_2)$, respectively.
When $\k_2 >0$ these expressions for the Casimirs cannot  of
course be interpreted in physical terms as energy and angular
momentum. 

We recall that each Lie algebra $g$ of table 1 admits three 
involutive automorphisms, which we will name according to their
natural interpretation in the kinematical case: parity
${\cal P}$, time-reversal ${\cal T}$ and their product  ${\cal
PT}$  defined by \cite{BLL}:
\be
\begin{array}{ll}
{\cal P}:&\quad (H,P_i,K_i,J)\to (  H, - P_i,-K_i,J)\cr
{\cal T}:&\quad(H,P_i,K_i,J)\to ( - H,  P_i,-K_i,J)\cr
{\cal PT}:&\quad(H,P_i,K_i,J)\to ( - H, - P_i,K_i,J) 
\end{array} 
\label{zb}
\ee
These mappings clearly leave the Lie brackets  (\ref{aa})
invariant. For further purposes we consider direct sum
decompositions of $g$ into anti-invariant and invariant generators
under the action of  the involutions 
${\cal PT}$ and ${\cal P}$:
\be
\begin{array}{llll}
{\cal PT}: \qquad &g=p^{(1)}\oplus h^{(1)}&\quad
p^{(1)}=\langle H,P_i\rangle &\quad
h^{(1)}=\langle  K_i,J\rangle \cr
{\cal P}: \qquad &g=p^{(2)}\oplus h^{(2)}&\quad
p^{(2)}=\langle P_i,K_i\rangle &\quad
h^{(2)}=\langle
H\rangle \oplus \langle J\rangle .
\end{array} 
\label{zc}
\ee
Both are Cartan decompositions, verifying
\be
[h^{(i)},h^{(i)}]\subset h^{(i)}\qquad [h^{(i)},p^{(i)}]\subset
p^{(i)}
\qquad [p^{(i)},p^{(i)}]\subset h^{(i)} .
\label{zd}  
\ee 
Notice that $h^{(i)}$ is always a Lie subalgebra of $g$, while
$p^{(i)}$ is only a subalgebra whenever  $\k_i=0$ ($i=1,2$), and
in that case it is  abelian: $[p^{(i)},p^{(i)}]=0$.
Hence $g$  is the Lie algebra of the motion
group $G$ of the following symmetrical
homogeneous spaces:
\be
\begin{array}{lll}
{\cal S}^{(1)}=G/H^{(1)}  &\quad
\mbox{dim}\,({\cal S}^{(1)})=3&\quad
\mbox{curv}\,({\cal S}^{(1)})=\k_1\cr
{\cal S}^{(2)}=G/H^{(2)}  &\quad
\mbox{dim}\,({\cal S}^{(2)})=4&\quad
\mbox{curv}\,({\cal S}^{(2)})=\k_2  
\end{array} 
\label{aag}
\ee
where $H^{(1)}$, $H^{(2)}$, the subgroups 
whose corresponding Lie algebras are $h^{(1)}$, $h^{(2)}$,
are the isotropy subgroups of a point/event  and a
(time-like) line, respectively. Therefore   ${\cal
S}^{(1)}$ is identified either to a three-dimensional  space of
points  or to a (2+1)-dimensional space-time, in both cases  with
constant curvature $\k_1$. Likewise, ${\cal S}^{(2)}$ is a
four-dimensional space, whose `points' can be identified to
(time-like) lines in the former space; this has a natural
connection and metric structure, whose  curvature turns out to be
`constant' (in some suitable rank-two sense which is  compatible
with the fact that this space always contains a flat submanifold 
whose dimension equals to the rank) and equals $\k_2$. 
 
To take the constant $\k_1$ (resp.\ $\k_2$)  equal to  zero is
equivalent to perform  an In\"on\"u--Wigner contraction
\cite{IW} starting from some algebra where $\k_1\neq 0$ (resp.\ 
$\k_2\neq 0$). In this contraction,  the invariant subalgebra is
$h^{(1)}$  (resp.\  $h^{(2)}$), while  the remaining  generators
are multiplied by a parameter
$\varepsilon$; the contracted algebra appears as the limit 
$\varepsilon\to 0$. If we perform this limiting procedure 
starting from the generic algebra (\ref{aa}) we find that a 
space-time contraction makes to vanish  the curvature
$\k_1$ of 
${\cal S}^{(1)}$ ($R\to \infty$), while a speed-space 
contraction makes zero the curvature $\k_2$ of  ${\cal S}^{(2)}$
(in the kinematical case $c\to
\infty$): 
\be
\begin{array}{llll}
\k_1\to 0:&\mbox{Space-time contraction}&\ 
(H,P_i,K_i,J)\to (\varepsilon H,\varepsilon P_i,K_i,J)&
\quad \varepsilon\to 0\cr
\k_2\to 0:&\mbox{Speed-space contraction}&\ 
(H,P_i,K_i,J)\to ( H,\varepsilon P_i, \varepsilon K_i,J) &
\quad \varepsilon\to 0 .
\end{array} 
\label{aah}
\ee

In table  1 horizontal arrows correspond to space-time
contractions and the vertical ones to  speed-space contractions.
The Lie algebra expansions we are going to describe in the next
sections are somewhat the opposite process and allow to recover
these constants starting from a contracted algebra.
 In geometrical terms, an expansion allows to introduce
curvature out of a flat space.


\sect{An expansion method}

Let $g$ and $g'$  be two  Lie algebras with commutation rules
given by (\ref{aa}). We suppose that
$g$ is a contracted algebra obtained  from $g'$ by making zero
{\em one} of the two constants $\k_1$ or $\k_2$,  say $\k_a$, so
that  $g'\to g$ when $\k_a\to 0$, while the other  constant, (say
$\k_b$) does not change.  Now we want  to consider the {\em
opposite\/} situation: we look to $g$ as the initial algebra, and
we aim to recover $g'$, which we shall call the {\em expanded}
algebra,  starting from
$g$;  for that we have to introduce a non-zero value for 
$\k_a$ in some way. In the sequel we explain   the expansion
method we propose. The index $a$ will always  refer to the
constant which is being `expanded' from a zero value to a
non-zero one in the expansion process.

Let ${\cal C}_1$, ${\cal C}_2$   the two Casimirs of the  initial
Lie algebra $g$ (with $\k_a=0$) and ${\cal C}'_1$, ${\cal C}'_2$
those of the final algebra $g'$ (with $\k_a\neq 0$ and the same
remaining constant $\k_b$ as $g$). A glance to the explicit
expressions (\ref{ab}) clearly shows two facts: a) ${\cal
C}_i=\left.{\cal C}'_i\right|_{\k_a=0}$, and b)
${\cal C}'_i$ is {\em linear} in the chosen $\k_a$.   This
suggests to split each `expanded' Casimir into two  terms
according to the presence of the constant $\k_a$. Obviously, the
term independent of $\k_a$ is just the `contracted' Casimir, so
these decompositions define, out of the formal expressions for
the initial and the expanded Casimirs, some elements  in the
universal enveloping algebra of the initial algebra as:
\be
{\cal C}'_1 = \ppa +\k_a \jja\qquad
{\cal C}'_2 = \ppb +\k_a \jjb   
\label{ba}
\ee
where $\k_a$ does not appear in any of the terms $\ppp_l$,
$\jjj_l$ ($l=1,2$).  We now consider the linear combination
\be
\jjj=\AA_1 \jja + \AA_2\jjb
\label{bb}
\ee
where $\AA_1$, $\AA_2$ are two constants to be determined and we
will assume we are working in the universal enveloping algebra of
$g$ within an irreducible representation of $g$. 

We define some elements in this universal enveloping algebra  as 
the following functions of the generators $X_k$ of $g$:
\be
X'_k:=\left\{\begin{array}{ll}
X_k&\quad \mbox{if}\quad [\jjj,X_k] =0  \cr
[\jjj,X_k]&\quad \mbox{if}\quad [\jjj,X_k]
\ne 0 
\end{array}
\right. .
\label{bc}
\ee
The aim is to make these elements $X'_k$ close a Lie  algebra
isomorphic to $g'$. Once $\jjj$ is given,  the commutators of
$X'_k$ are completely determined, so that the only freedom at our
disposal in this procedure lies in the choice of the constants
$\AA_l$. 

The computations of commutators of the new elements $X'_k$ can be
shortcut in some cases by use of the following result:  
 
\noindent
{\bf Proposition 1.} Suppose that the initial Lie algebra   $g$
with generators $X_k$ has a direct sum   decomposition as vector
space as  $g=\p\oplus \h$ where $\h$ is the subalgebra determined
by the condition $\h=\langle X_k\  |\   [\jjj,X_k] =0\rangle$ and
$\p$ is some vector subspace supplementary to $\h$ (notice that
all elements in $\p$ do not commute with $\jjj$). Suppose also
that for commutators of elements in
$\h$ and $\p$ we have:
\be
[\h,\h]\subset \h \qquad [\h,\p]\subset \p  .
\label{bcd}
\ee
Then the generators $X'_k$ defined by (\ref{bc}) for the
expanded algebra $g'$ have the `same' Lie brackets 
$[\h',\h']$ and $[\h',\p']$ as the
initial algebra $g$.

The proof is trivial for $[\h',\h']$ as the
generators involved are invariant
  in the expansion and they directly span the   Lie subalgebra 
$\h'$. For $[\h',\p']$  we compute a generic Lie bracket 
between   $X'_l\in
\h'$ and $X'_m\in \p'$:
\be
[X'_l,X'_m]=[X_l,\jjj X_m - X_m  \jjj]=
\jjj [X_l,X_m]-[X_l,X_m]\jjj .
\ee
As $[\h,\p]\subset \p$, the commutator
$[X_l,X_m]=C_{lm}^n X_{n}\in \p$,  so that
\be
[X'_l,X'_m] = C_{lm}^n(\jjj X_{n} -  X_{n} \jjj) =
C_{lm}^n X'_n\in \p'.
\ee
We remark that the decomposition in Proposition 1 is defined in a
way independent to the Cartan decompositions in (\ref{zd}),  but
it might coincide with them.
In any case, the aim of the expansion idea is to get the
commutation relations of the Lie algebra $g'$ for the  new
generators $X'_k$.  Whenever the hypotheses of the proposition 1
are fulfilled, part of these commutation relations are
automatically satisfied and to get the correct expanded
commutation relations  we only have to compute the brackets
$[\p',\p']$ and to enforce for them the corresponding  
commutation relations of $g'$. In this way we obtain  some
equations involving $\AA_l$, $\ppp_l$ and $\k_a$; their solutions
characterize the constants $\AA_l$. The coefficient $\k_a$ (not
appearing in $g$) is introduced in this last step. 

In the next sections we apply this method to the algebras of table 1,
reversing the direction of the contraction arrows. As we have two
`curvatures' we will consider two types of expansions: space-time
expansions, which out of $\k_1=0$  recover $\k_1$,  and
speed-space expansions which similarly recover $\k_2$.  In most
cases the assumptions of the proposition 1 will be satisfied,
and  for each expansion  starting from $\k_a=0$ we will find that
initially $[\p,\p]=0$, and after the expansion $[\p',\p']\subset
\h'$ due to the presence of an `expanded' non-zero value for
$\k_a$.  It is remarkable that in the expansion  which goes from
Galilei to Poincar\'e it will be necessary to  consider the
initial Galilei algebra with a central extension; however, the
procedure just described is still valid. Actually, this fact
already happens in   1+1 dimensions \cite{nieto}.


\sect{Space-time expansions or $\k_1$-expansions}

The purpose of this section is to discuss  the expansions which
starting from the algebra with   $\k_1=0$ `introduce' a non-zero
value  for the constant $\k_1$. The value of $\k_2$ will remain
unchanged in the expansion. Some details are slightly different
according to either $\k_2\neq 0$ or $\k_2= 0$, so we will present
these two cases separately. When applied to the kinematical
algebras, this expansion leads from the Galilei algebra to the two
Newton-Hooke ones, and from the  Poincar\'e case to the two
de Sitter algebras. In the non-kinematical case
$\k_2>0$ the expansion carries from 3d Euclidean algebra  to
either the elliptic or hyperbolic ones. 

\subsect{From Poincar\'e to de Sitter}

We consider as initial algebras those with $\k_1=0$ and $\k_2\ne
0$ which are  the  Euclidean  $iso(3)$  (for $\k_2>0$)  and
the   Poincar\'e $iso(2,1)$  (for $\k_2<0$) algebras; they
 are  associated to a flat 3d Euclidean space and to  a
relativistic flat (2+1)d space-time, respectively. The  Lie
brackets which due to the initial condition
$\k_a
\equiv
\k_1=0$ vanish in the general commutation relations  (\ref{aa})
are:
\be
[P_1,P_2]=0\qquad 
[H, P_i]=0, 
\label{ca}
\ee
and the two Casimirs  (\ref{ab}) reduce to
\be
{\cal C}_1=\k_2 H^2+ P_1^2+P_2^2 
\qquad
{\cal C}_2=\k_2HJ-P_1 K_2 + P_2 K_1 .
\label{cb}
\ee

The expansion to the  $so(3)$ or $so(2,1)$ algebras which
correspond to $\k_1 \neq 0$ and the same initial value for $\k_2$
requires to replace  the  three Lie brackets in (\ref{ca})  by
those corresponding to
$\k_1\neq 0$, which read:
\be
[P'_1,P'_2]=\k_1\k_2 J'\qquad 
[H', P'_i]=\k_1 K'_i .
\label{canew}
\ee
We  split the 
Casimirs of the final semisimple algebras as:
\be 
\begin{array}{ll}
\expandcasa = \ppa +\k_1 \jja\qquad  &\jja= K_1^2+ K_2^2+
\k_2J^2  \cr
\expandcasb = \ppb \qquad &\jjb=0 .
\end{array}
\label{cc}
\ee
Hence the linear combination (\ref{bb})  has a single term:
$\jjj=\AA_1\jja$. The new generators coming from  
(\ref{bc}) read:
\be
\begin{array}{l}
 \expandk_1=K_1\qquad \expandk_2=K_2 \qquad
\expandj=J \cr
 \expandh= 2
\AA_1 (K_1 P_1+ K_2 P_2+ \k_2 H) \cr
\expandp_1=2 \k_2
\AA_1 (J P_2- K_1 H+ P_1 ) \cr
\expandp_2=  2 \k_2 
\AA_1 (-J P_1- K_2 H +P_2) .
\end{array}
\label{ce}
\ee 
In this case, the decomposition  $g=\p\oplus\h$  coincide with
the Cartan decomposition $g= p^{(1)}\oplus h^{(1)}$, and  the
three generators which are unchanged by the expansion  close the
isotropy subalgebra of a point/event  $h^{(1)}$ (\ref{zc}).
Taking into account (\ref{zd}) it is clear that proposition 1 can
be applied. The expansion depends on a single parameter $\AA_1$,
whose value (if the expansion indeed exists) is obtained by
enforcing (\ref{canew}) for the three commutators
$[\expandp_1,\expandp_2]$, $[\expandh,\expandp_i]$.  Let us
compute, for instance,
\be
\begin{array}{l}
[\expandh,\expandp_1]= 
4\k_2\AA_1^2(-\k_2 K_1 H^2- K_1 P_1^2
- K_1 P_2^2)\cr
 \qquad\qquad
=-4\k_2\ \AA_1^2 K_1(\k_2 H^2+ P_1^2+P_2^2) 
=-4\k_2 \AA_1^2\expandk_1 \ppa .
\end{array}
\label{cf}
\ee 
Remark the automatic appearance of the Casimir $\ppa$;  this will
happen in all expansions we will deal with.  Since the commutator
must be equal to $\k_1\expandk_1$ we have
\be
\AA_1^2=-\frac{\k_1}{4\k_2\ppa} .
\label{cg}
\ee
It can be checked that the two remaining Lie brackets lead to the
same condition.

Note that $\AA_1$ is not strictly speaking a  number, but depends
on the generators of the initial algebra {\em only} through the
Casimir $\ppa$. Within any irreducible representation  of the
initial algebra, $\AA_1$ turns into a scalar value.

According to the different values for the initial  constant
$\k_2\neq 0$ (remind we start from $\k_1=0$) and  the possible
choices of the expansion parameter $\AA_1$ (i.e., of the final
$\k_1$), the process just described leads to the algebras
displayed in the diagram:

\medskip

\begin{tabular}{lccccc}
&$so(4)$&$\longleftarrow$& $iso(3)$&$\longrightarrow$& $so(3,1)$\\
$\k_2>0$&$(+,+)$& &$(0,+)$& &$(-,+)$\\
&Elliptic& &Euclidean& &Hyperbolic\\[0.6cm]
&$so(2,2)$&$\longleftarrow$& $iso(2,1)$&$\longrightarrow$&
$so(3,1)$\\ 
$\k_2<0$&$(+,-)$& &$(0,-)$& &$(-,-)$\\
&Anti-de Sitter& &Poincar\'e& &de Sitter
\end{tabular} 

\medskip

This type of expansions allows  us to `recover' a space of 
constant curvature (elliptic/hyperbolic, or anti de Sitter/de
Sitter) out of a flat space, either the  3d
Euclidean space or the (2+1)d Minkowskian space-time.

\subsect{From extended Galilei to Newton--Hooke}

In the non-generic case $\k_2=0$, we must start  the
$\k_1$-expansion from the degenerate Galilei algebra. We want to
keep $\k_2=0$ but to introduce  $\k_1\ne 0$,  then reaching the
Newton--Hooke algebras. The commutators which are zero in the
initial algebra but not in the expanded one are only: 
\be
[H,P_i]=0 .
\label{fa}
\ee
The Galilean Casimirs  read
\be
{\cal C}_1=P_1^2+P_2^2\qquad
{\cal C}_2= -P_1 K_2+ P_2 K_1 .
\label{fb}
\ee
We split  the Newton--Hooke invariants   as 
\be 
\begin{array}{ll}
\expandcasa=\ppa + \k_1 \jja \qquad  & \jja =  K_1^2+K_2^2 \cr
\expandcasb=\ppb  \qquad & \jjb=0 .
\end{array}
\label{fc}
\ee 
Thus $\jjj=\AA_1\jja$.  Should we apply blindly the expansion
recipe, from (\ref{bc}) we obtain  that
$\expandh=2\AA_1( K_1 P_1+K_2 P_2 )$, all other generators being
unchanged.  Although   proposition 1 cannot be used in this case
to shortcut computations (note that  $[\h,\p]\subset \h$),  it
can be checked that the new generators so obtained do  close a
Lie algebra, which is however {\em not} within the set of the
algebras described in (\ref{aa}). In this case the initial Lie
algebra is too much contracted (or abelian) to be able to act as
a germ for an expansion to the Newton--Hooke algebras. However
this problem can be circumvented in the same way as in the
(1+1)-dimensional case
\cite{nieto}:  starting not from Galilei algebra itself,  but
from a central extension, with central generator
$\Xi$ and characterized by a parameter  $\ext$, the mass  of a
free particle. The  Lie brackets of this  extended Galilei
algebra  are given by
\be
\begin{array}{lll}
[J,P_i]=\epsilon_{ij} P_j&\qquad [J,K_i]=\epsilon_{ij} K_j&
\qquad [J,H]=0\cr
[P_1,P_2]=0&\qquad [K_1,K_2]=0 &
\qquad [P_i,K_j]=\delta_{ij}\ext\Xi\cr
[H,P_i]=0&\qquad [H,K_i]=-P_i &\qquad [\Xi,\,\cdot\,]=0 .
\end{array}
\label{fd}
\ee
 We keep  $\jjj=\AA_1 ( K_1^2+K_2^2)$ and apply again the recipe
(\ref{bc}) to define the expanded generators;  due to the presence
of the central extension the results found formerly  change, and
now we get:
\be
\begin{array}{l}
 \expandk_1=K_1\qquad \expandk_2=K_2 \qquad
\expandj=J \cr
\expandh=2 \AA_1 (K_1 P_1 + 
K_2 P_2 + \ext\Xi)\cr
\expandp_1= -2\AA_1 \ext\Xi K_1 \qquad
\expandp_2= -2\AA_1   \ext\Xi K_2 .
\end{array}
\label{fe}
\ee 
Hence the subalgebra $k$ unchanged by the expansion  coincides
with $h^{(1)}$, the isotropy subalgebra of an event. In spite of
the central extension, the same reasonings of the proposition 1
show that  the Lie brackets $[\h',\h']$ and $[\h',\p']$ are kept
in same form as in the non-extended initial Galilei algebra.  The
remaining commutators $[\p',\p']$ lead to 
\be
[\expandh,\expandp_i]=-4\AA_1^2  \ext^2 \Xi^2   K_i\equiv
\k_1\expandk_i \qquad [\expandp_1,\expandp_2]=0,
\label{ff}
\ee 
and consequently
\be
\AA_1^2=-\frac{\k_1}{4  \ext^2 \Xi^2} .
\ee

This Galilean expansion recovers a non-zero 
curvature $\k_1$ out of the flat Galilei space-time,  while
keeping $\k_2=0$ which is accompanied by the presence  of
`absolute time', producing the two curved `absolute time'
Newton-Hooke space-times and thereby completing the non-generic
missing middle line in the diagram of section 4.1:

\medskip

\noindent
\begin{tabular}{lccccc}
&$t_4(so(2)\oplus
so(2))$&$\longleftarrow$&
${\overline
{iiso}}(1,1)$&$\longrightarrow$&$t_4(so(2)\oplus
so(1,1))$\\ 
$\k_2=0$&$(+,0)$& &$(0,0)$& &$(-,0)$\\
&Oscillating NH& &Extended Galilei& &Expanding NH
\end{tabular} 


\sect{Speed-space expansions or $\k_2$-expansions}

In this section, we switch roles for $\k_1$ and $\k_2$,  and we
discuss expansions which starting from the algebra with  
$\k_2=0$ `introduce' a non-zero value for the  constant $\k_2$,
the value of $\k_1$ being  unchanged. Again some details are
slightly different according to either $\k_1=0$ or $\k_1\neq 0$,
so we will study these separately. The name speed-space we give
to these expansions is justified because when applied to the
kinematical algebras, these expansions lead from the Galilei
algebra to the Poincar\'e or to the 3d Euclidean one, while from
Newton-Hooke the expansion leads either to the two de Sitter
algebras, or to the 3d elliptic and hyperbolic algebras. 

\subsect{From Newton--Hooke to de Sitter}

We consider as the initial algebras those with  $\k_2=0$ and a
fixed $\k_1\neq 0$, that is,  the   Newton--Hooke ones. There are
four  Lie brackets of (\ref{aa}) which are zero in the initial
algebra but should be different from zero in the expanded one:
\be
[P_1,P_2]=0\qquad 
[K_1, K_2]=0\qquad
[P_1, K_1]=0  \qquad
[P_2, K_2]=0.
\label{da}
\ee
The two Casimirs   (\ref{ab}) are now
\be
{\cal C}_1= P_1^2+P_2^2+
\k_1(K_1^2+ K_2^2)  \qquad
{\cal C}_2=  -P_1 K_2+ P_2 K_1  .
\label{db}
\ee
  We decompose the two Casimir invariants of the 
algebras we want to reach by expansion  (isomorphic to either
$so(3)$ or $so(2,1)$) by taking into account the expansion
constant $\k_2$:
\bea
&&\expandcasa=\ppa + \k_2 \jja \qquad
\jja = H^2+ \k_1J^2\cr
&& \expandcasb=\ppb + \k_2
\jjb \qquad 
\jjb =J H .
\label{dd}
\eea
Therefore the element (\ref{bb}) has two  terms and gives rise to
the new generators defined by
\be
\begin{array}{l}
\expandh=H \qquad \expandj=J \cr
\expandp_1=2\k_1 \AA_1 ( K_1 H +  J P_2)
+\AA_2 (P_2 H+\k_1 J K_1) \cr
\expandp_2=2\k_1 \AA_1
(  K_2 H- J P_1) +\AA_2
(-P_1 H+\k_1 J K_2) \cr
\expandk_1=2\AA_1 (- P_1 H +\k_1 J K_2)
+\AA_2 (K_2 H - J P_1) \cr
\expandk_2=2\AA_1 (- P_2 H -\k_1 J K_1)
+\AA_2 (- K_1 H - J P_2).
\end{array}
\label{de}
\ee 

In this case the  decomposition  $g=\p\oplus \h$ coincides
with  the Cartan one associated to the involution $\cal P$, and
the invariant generators   $H$ and $J$ generate  the isotropy
subalgebra $h^{(2)}$ of  a (time-like) line (\ref{zc}). This means
that proposition 1 can be applied. Thus we have only to compute
the Lie brackets  involving the  four generators $\expandp_i$,
$\expandk_i$. Let us choose, for instance,
\bea
[\expandp_1,\expandp_2]\!\!&=&\!\!
 -4 \k_1^2  \AA_1^2  (2 K_1 P_2 H -2 K_2   P_1 H\cr
&&\qquad   + J P_1^2
+J P_2^2+\k_1J K_1^2+\k_1J K_2^2)\cr
&&-  \k_1   \AA_2^2  ( 2 K_1 P_2 H -2 K_2   P_1 H\cr
&&\qquad  + J P_1^2
+J P_2^2+\k_1J K_1^2+\k_1J K_2^2)\cr
&&- 2\k_1   \AA_1  \AA_2   (2  P_1^2 H + 2  P_2^2 H
+2\k_1 K_1^2 H + 2\k_1 K_2^2 H\cr
&&\qquad +4 \k_1 J K_1 P_2
-4\k_1 J K_2 P_1) .
\label{df}
 \eea
We introduce in this expression the the Newton--Hooke  Casimirs
(\ref{db}) and we get:
\bea
[\expandp_1,\expandp_2]\!\!&=&\!\! -4 \k_1^2  \AA_1^2  (2 
\ppb H + J\ppa) -  \k_1   \AA_2^2  (2  \ppb H+
J \ppa)\cr 
&&- 2\k_1  \AA_1  \AA_2  (2
\ppa H +4\k_1 J\ppb )\cr
\!\!&=&\!\! -(8\k_1^2 \ppb  \AA_1^2 +2\k_1 \ppb 
\AA_2^2  +  4
\k_1 \ppa \AA_1 \AA_2)\expandh \cr
&&-(4\k_1^2 \ppa  \AA_1^2 +\k_1\ppa  \AA_2^2  +  8
\k_1^2  \ppb \AA_1 \AA_2)\expandj   
\label{dg}
\eea
and by imposing (\ref{dg}) to be equal to  $\k_1\k_2\expandj$ we
get  two quadratic equations in the constants $\AA_l$:
\be
\begin{array}{l}
4\k_1 \ppa  \AA_1^2 +\ppa  \AA_2^2  +  8 \k_1 
\ppb \AA_1 \AA_2=-\k_2\cr
4\k_1 \ppb  \AA_1^2 +\ppb  \AA_2^2  +  2
\ppa \AA_1 \AA_2=0 .
\end{array}
\label{dh}
\ee 
If we calculate any other Lie bracket   (\ref{da}) with
the new generators (\ref{de}) we 
obtain the same equations (\ref{dh}). Moreover, we have also to
compute the commutators $[\expandp_1,\expandk_2]$ and
$[\expandp_2,\expandk_1]$; they are directly zero and do not
originate any relation for the constants $\AA_l$.

Hence, within an irreducible representation of the  initial
algebra, the Casimirs appear replaced by their eigenvalues, and
the solutions in  $\AA_1$ and $\AA_2$ for
the quadratic equations (\ref{dh}) afford the expansions we are
looking for. 

These expansions  which start from the Newton--Hooke algebras 
introduce  the constant $\k_2$ in the four-dimensional  spaces of
lines ${\cal S}^{(2)}$  ($\k_2=-1/c^2$ when it is negative), thus
eliminating the `absolute time' character and giving rise to the
curved relativistic de Sitter algebras; they embrace the
following cases:

\begin{center}
\begin{tabular}{ccccc}
$\k_1>0$&&&&$\k_1<0$\\[0.3cm]
$so(4)$&&&&$so(3,1)$\\
$(+,+)$&&&&$(-,+)$\\
Elliptic&&&&Hyperbolic\\ 
$\uparrow$&&&&$\uparrow$\\
$t_4(so(2)\oplus so(2))$&&&&$t_4(so(2)\oplus
so(1,1))$\\
$(+,0)$&&&&$(-,0)$\\
Oscillating NH&&&&Expanding NH\\
$\downarrow$&&&&$\downarrow$\\
$so(2,2)$&&&&$so(3,1)$\\
$(+,-)$&&&&$(-,-)$\\
 Anti-de Sitter&&&&de Sitter 
\end{tabular} 
\end{center}

\subsect{From Galilei to Poincar\'e}

Finally we consider the $\k_2$-expansion starting  from the
Galilei algebra which has not only $\k_2=0$ but also $\k_1=0$. We
want to obtain Lie algebras with $\k_2\ne 0$, but keeping the
Galilean value of $\k_1$. The Lie brackets that we have to make
different from  zero   read (see (\ref{aa})):
\be
[K_1,K_2]=0\qquad
[P_1,K_1]=0\qquad
[P_2,K_2]=0 .
\label{ea}
\ee
By taking into account the Galilean Casimirs (\ref{fb})  we write
the invariants  (\ref{ab}) with $\k_1=0$ as  
\bea
&&\expandcasa=\ppa + \k_2 \jja \qquad \jja =   H^2\cr
&&  \expandcasb= \ppb + \k_2
\jjb \qquad 
\jjb =JH .
\label{ed}
\eea
Therefore the generators for the expanded algebras are
\be
\begin{array}{l}
\expandh= H \qquad \expandj = J\cr
\expandp_1  = \AA_2 P_2 H \qquad
\expandp_2=  -\AA_2  P_1 H \cr
\expandk_1=- 2\AA_1 P_1 H
+\AA_2 (K_2 H - J P_1) \cr
\expandk_2=-2\AA_1 P_2 H
- \AA_2 ( K_1 H + J P_2).
\end{array}
\label{ee}
\ee 
As in the previous expansion, we have only to compute  the
commutators between the generators $\expandp_i$,  $\expandk_i$.
Enforcing the values they should have in the expanded
algebra we get the constants $\AA_l$:
\be
\begin{array}{l}
[\expandp_i,\expandk_i]=-\AA_2^2(P_1^2 H +P_2^2 H)=
-\AA_2^2 \ppa H\equiv \k_2\expandh \cr
[\expandk_1,\expandk_2]=
-\AA_2^2(2 K_1 P_2 H - 2 K_2 P_1 H + J P_1^2 +J P_2^2)
-4\AA_1\AA_2 (P_1^2  +P_2^2) H\cr
\qquad\qquad
=-2\AA_2 ( 2 \AA_1  \ppa + \AA_2\ppb) H
-\AA_2^2 \ppa J\equiv
\k_2 \expandj  \cr
[\expandp_1,\expandp_2]=0\qquad 
[\expandp_1,\expandk_2]=0\qquad 
[\expandp_2,\expandk_1]=0,
\end{array}
\label{ef}
\ee 
that is,
\be
\AA_2^2=-\frac{\k_2}{\ppa}\qquad \AA_1=-\frac{\AA_2\ppb}{2\ppa} .
\label{ei}
\ee

This Galilean expansion   which recovers the curvature $\k_2$
of the space of (time-like) lines ${\cal S}^{(2)}$  gives rise to
the Euclidean and Poincar\'e algebras, and involve the eigenvalue
of both Casimirs:

\begin{center}
\begin{tabular}{c}
$\k_1=0$\\[0.3cm]
$iso(3)$\\
$(0,+)$\\
Euclidean\\
$\uparrow$\\
$iiso(2)$\\
$(0,0)$\\
Galilei\\
$\downarrow$\\
$iso(2,1)$\\
$(0,-)$\\
Poincar\'e
\end{tabular} 
\end{center}

\newpage


\sect{Concluding remarks}

We have presented an expansion method which 
allows to reverse all the contraction arrows of the Lie algebras
displayed in table 1. We would like to  stress several points
which turn out to be relevant, and which may hint towards the
still rather unknown extension to the expansion procedure to
either higher dimensional situations or to higher rank cases. 

First, in the $\k_1$-expansions recovering the curvature
$\k_1$ of the space  ${\cal S}^{(1)}$ only the first  Casimir
${\cal C}_1$ appears, while both Casimirs participate in the
$\k_2$-expansions making different from zero the curvature $\k_2$
of the space of lines ${\cal S}^{(2)}$.  As the rank of the
homogeneous spaces ${\cal S}^{(1)}$,
${\cal S}^{(2)}$ is one and two, respectively, the
results obtained  seem to confirm the expected relationship
between the rank of the space and the number of Casimirs needed
to perform the expansion. This idea is in agreement with the known  
generalizations to arbitrary dimension and variants of expansions  
$iso(p,q)\to so(p,q)$ (associated to  ${\cal S}^{(1)}$)
\cite{Gilmore,Rosen} which only involve a single  Casimir (the
quadratic one) and are in a sense direct generalizations to any
higher dimension from the $\k_1$-expansions we discuss here. 

Second, the role of extended algebras as the starting  point for
the expansion needs also clarification. This role clearly depends
on the type of expansion to be done. While the starting point for
the $\k_1$-expansion of Galilei algebra should be an extended
Galilei algebra, this is not necessary for the   $\k_2$-expansion
of the same algebra. A complete and systematic study of all the
central extensions of the quasi-orthogonal algebras is available
\cite{extOrtog}, and should be the starting point  to understand
the role these extensions play in the expansion process, a
problem which  deserves further study. 

But the more interesting open question  would  be to know whether
or not some suitable `extension' of the method we have proposed is still
applicable for higher dimensions.  It is natural to suppose that
whatever the correct method should be, it should rest again on
the Casimirs of the initial and the expanded algebra, and their
dependence on the expansion constant $\k_a$. Two facts will
likely complicate the issue under discussion. First, further to
the quadratic Casimir, the additional ones are higher order (this
is masked in the $so(4)$ family because the additional Casimir
here is a perfect square and it can be considered as an extra
quadratic one). Second, the dependence of higher order Casimirs
on the expansion constant $\k_a$ is also known in the general
case \cite{casimir} and this dependence is not only linear but
also given by a higher order polynomial. The analysis of the next
situation,  the (3+1)-dimensional case, would help to clarify the
above questions.


\bigskip\bigskip

\noindent
{\Large{{\bf Acknowledgments}}}

\bigskip

 This work was partially supported by DGES (Project 
94--1115) from the Ministerio de Educaci\'on y Cultura  de
Espa\~na and by Junta de Castilla y Le\'on (Projects CO1/196  
and CO2/197).

\bigskip


\end{document}